\documentclass[journal]{IEEEtran}
\usepackage{amsmath,amsfonts}
\usepackage{array}
\usepackage[caption=false,font=normalsize,labelfont=sf,textfont=sf]{subfig}
\usepackage{textcomp}
\usepackage{stfloats}
\usepackage{url}
\usepackage{verbatim}
\usepackage{graphicx}
\usepackage{cite}
\usepackage[linesnumbered,ruled,vlined]{algorithm2e}
\usepackage{tabularx}
\usepackage{adjustbox}
\usepackage{booktabs}
\usepackage{threeparttable}
\usepackage{makecell}
\usepackage{enumitem}
\usepackage{orcidlink}

\usepackage{pifont}
\usepackage[table]{xcolor}

\usepackage[most]{tcolorbox}
\newtcolorbox{mybox}{
  colback=gray!10,     % 背景颜色（稍深的浅灰）
  colframe=black!75,   % 边框颜色（稍深的深灰）
  boxrule=0.5pt,       % 边框线宽
  arc=2pt,             % 圆角
  outer arc=2pt,
  left=6pt,            % 左内边距
  right=6pt,           % 右内边距
  top=6pt,             % 上内边距
  bottom=6pt,          % 下内边距
  boxsep=0pt,          % 内部内容与边框之间的额外间距
}

\usepackage{hyperref}
\hypersetup{
    colorlinks=true,   % true: 链接文字有颜色；false: 链接有彩色框
    linkcolor=black,   % 文本链接颜色（章节、公式、图表引用等）
    citecolor=black,    % 引用文献颜色
    urlcolor=black      % 超链接 URL 的颜色
}

\begin{document}

\title{Enhancing LLM-Based Bug Reproduction for Android Apps via Pre-Assessment of Visual Effects}

\author{Xiangyang Xiao~\orcidlink{0009-0006-0352-6739},
Huaxun Huang$^{*}$~\orcidlink{0000-0002-1778-3721},
Rongxin Wu~\orcidlink{0000-0002-4648-3795}
\thanks{The authors are with Xiamen University, Xiamen, Fujian, China (e-mail: xiangyangxiao@stu.xmu.edu.cn; huanghuaxun@xmu.edu.cn; wurongxin@xmu.edu.cn). Huaxun Huang is the corresponding author.}}

% \author{IEEE Publication Technology,~\IEEEmembership{Staff,~IEEE,}
        % <-this % stops a space

% The paper headers
% \markboth{Journal of \LaTeX\ Class Files,~Vol.~14, No.~8, August~2021}%
% {Shell \MakeLowercase{\textit{et al.}}: A Sample Article Using IEEEtran.cls for IEEE Journals}

% \IEEEpubid{0000--0000/00\$00.00~\copyright~2021 IEEE}
% Remember, if you use this you must call \IEEEpubidadjcol in the second
% column for its text to clear the IEEEpubid mark.

\maketitle

\begin{abstract}
    In the development and maintenance of Android apps, the quick and accurate reproduction of user-reported bugs is crucial to ensure application quality and improve user satisfaction. However, this process is often time-consuming and complex. Therefore, there is a need for an automated approach that can explore the Application Under Test (AUT) and identify the correct sequence of User Interface (UI) actions required to reproduce a bug, given only a complete bug report. Large Language Models (LLMs) have shown remarkable capabilities in understanding textual and visual semantics, making them a promising tool for planning UI actions. Nevertheless, our study shows that even when using state-of-the-art LLM-based approaches, these methods still struggle to follow detailed bug reproduction instructions and replan based on new information, due to their inability to accurately predict and interpret the visual effects of UI components. To address these limitations, we propose LTGDroid. Our insight is to execute all possible UI actions on the current UI page during exploration, record their corresponding visual effects, and leverage these visual cues to guide the LLM in selecting UI actions that are likely to reproduce the bug. We evaluated LTGDroid, instantiated with GPT-4.1, on a benchmark consisting of 75 bug reports from 45 popular Android apps. The results show that LTGDroid achieves a reproduction success rate of 87.51\%, improving over the state-of-the-art baselines by 49.16\% and 556.30\%, while requiring an average of 20.45 minutes and approximately \$0.27 to successfully reproduce a bug. The LTGDroid implementation is publicly available at \href{https://github.com/N3onFlux/LTGDroid}{https://github.com/N3onFlux/LTGDroid}.
\end{abstract}


\section{Introduction}

\IEEEPARstart{A}{ndroid} apps have become an indispensable part of modern life.
Debugging and fixing bugs are critical tasks in Android app development. Studies have shown that 88\% of users may stop using an app when they experience repeated bugs~\cite{APPLAUSE}, highlighting the need to quickly resolve bugs in Android apps to maintain user satisfaction. Bug reporting systems, such as GitHub Issue Tracker~\cite{githubissuetracker} and Google Code~\cite{googlecode}, are widely used to collect and manage bug reports. These reports typically describe both the observed and expected behaviors, and often provide steps to reproduce (S2Rs) the bugs, which assist developers in reproducing and resolving the issues. However, these reports are often written in informal natural language, which can be imprecise and incomplete, making it difficult for developers to reproduce the reported bugs. Even with well-documented reports, reproducing bugs can be challenging due to the complex, event-driven UIs of Android apps. These challenges make the bug reproduction process time-consuming and error-prone, reducing its effectiveness. Automated bug reproduction techniques can assist developers in quickly reproducing bugs, thereby improving repair efficiency and software quality.

Various approaches have been proposed for automated bug reproduction~\cite{Yakusu, ReCDroid, ReCDroid+, ReproBot, ScopeDroid, Roam, BugSpot, AdbGPT, ReBL}. A class of methods~\cite{Yakusu, ReCDroid, ReCDroid+, ReproBot, Roam} relies on traditional Natural Language Processing (NLP) techniques to extract steps to reproduce (S2R) from bug reports and map them to user interface (UI) widgets of the target app. Most of these approaches are based on identifying specific grammar patterns or analyzing the linguistic structure of the bug report to extract relevant actions and entities. However, such reliance on predefined or heuristic-based language patterns constrains their ability to understand the full semantic richness and diversity of user-written S2R instructions. Consequently, these approaches may struggle to accurately interpret bug reports whose descriptions fall outside commonly observed patterns, resulting in difficulty in bridging the semantic gap between natural language bug reports and the actual UI context of the app. Recently, Large Language Models (LLMs) have demonstrated capabilities in understanding the semantics of text, code, and images. Some studies~\cite{AdbGPT, ReBL} have attempted to use LLMs to guide LLM-based agents in generating UI actions by combining contextual information such as the S2R, the current UI screen, and the reproduction history.
However, approaches that rely solely on the above contextual information are still limited in their ability to predict the outcomes of UI actions. As a result, incorrect UI actions may be generated, causing the reproduction process to deviate from the correct steps. This deviation is fundamentally due to the inability to predict the consequences of each action.
In our ablation study (see Section~\ref{sec:RQ2}), compared with the ablated version without the ability to predict the consequences of UI actions, our approach increased the overall reproduction success rate from 51.11\% to 87.51\%, corresponding to a relative improvement of 67.31\%.

To address these challenges, we propose LTGDroid, an LLM-based approach for generating UI actions for Android apps based on bug reports. Unlike existing methods that rely on LLMs to directly plan the next UI action based on screen semantics, LTGDroid enumerates all available UI actions on each UI screen and pre-assesses their resulting runtime behaviors. LTGDroid then guides the LLM to select UI actions that align with the S2R in the bug reports. Although there is a potential concern that enumerating all possible actions may theoretically reduce efficiency in bug reproduction, our experimental results show that this pre-assessment procedure only leads to a limited increase in the number of UI actions while improving the overall reproduction success rate.

To evaluate the effectiveness of LTGDroid, we built a benchmark dataset comprising 75 bug reports drawn from 45 open-source Android apps. We compare LTGDroid with state-of-the-art LLM-based approaches, including ReBL~\cite{ReBL} and AdbGPT~\cite{AdbGPT}, all powered by GPT-4.1 for a fair comparison. The evaluation results show that LTGDroid achieves an overall success rate of 87.51\%, outperforming the baselines with relative improvements of 49.16\% and 556.30\%, respectively. Efficiency analysis further demonstrates that LTGDroid requires an average of 20.45 minutes and approximately \$0.27 to successfully reproduce a single bug.

In summary, the main contributions of this paper are as follows:
\begin{itemize}[leftmargin=*]
    \item To the best of our knowledge, we are the first to investigate leveraging runtime behaviors caused by UI actions to assist LLMs in automated bug reproduction.
    \item We implemented an open-source tool, LTGDroid, that effectively utilizes runtime behaviors for automated bug reproduction.
    \item We evaluate LTGDroid on real-world bug reports, demonstrating that it achieves a high success rate of bug reproduction within a reasonable number of UI actions.
\end{itemize}

\section{Background}

\subsection{Android UI Model}

In Android apps, a UI page can be represented as a tree-structured UI state $S$, where each node corresponds to a UI widget $w$. This tree-based representation follows the hierarchical layout mechanism of Android, where a root view contains multiple nested child views. Each widget $w$ is associated with a set of attributes (e.g., \texttt{android:text}, \texttt{android:id}) that determine both its visual appearance and its interaction semantics.

A UI action is formally defined as $a = (t, w, o)$, where $t$ denotes the action type (e.g., \texttt{Click}, \texttt{LongClick}, \texttt{InputText}, \texttt{Swipe}), $w$ represents the target widget on which the action is performed (which may be null, such as in a global \texttt{Swipe} action), and $o$ is an optional data field (e.g., the input text in an \texttt{InputText} action). Executing an action $a$ on a UI state $S$ results in a transition to a new state $S'$, written as $S \xrightarrow{a} S'$.
% For example, performing an \texttt{InputText} action on a text field leads to a new UI state in which the entered text is rendered on the screen. 
Since the view hierarchy does not always fully reflect the rendered visual state of an app (e.g., enabling night mode alters the appearance while the corresponding view hierarchy remains unchanged), we treat each action as leading to a new UI state regardless of whether the underlying view hierarchy changes.

Such action-induced transitions collectively form a UI state transition graph of the app, which provides the foundation for automated exploration. In the context of bug reproduction, modeling UI states and actions in this way enables systematic reasoning about how a sequence of user interactions can trigger and reproduce a reported bug.

\subsection{Large Language Models}

Large Language Models (LLMs) typically refer to Transformer-based models that contain billions of parameters and are trained on massive text corpora. Representative examples include ChatGPT~\cite{ChatGPT}, GPT-4~\cite{GPT-4}, PaLM~\cite{PaLM}, and LLaMA~\cite{LLaMA}. Beyond processing and understanding natural language, modern LLMs are increasingly capable of capturing the semantics of images. Through multimodal training, these models can jointly analyze and interpret textual and visual information, enabling them to extract meaningful signals from images and align them with textual descriptions. This cross-modal semantic understanding allows LLMs to reason about complex relationships between different data modalities. Such capabilities make LLMs a promising foundation for automating bug report reproduction, where accurate interpretation of both textual bug descriptions and visual UI contexts is essential.

Another important advancement is the use of Chain-of-Thought (CoT) reasoning~\cite{wei2022chain}. CoT enables LLMs to explicitly decompose complex tasks into intermediate reasoning steps, rather than producing direct answers in a single step. This mechanism has been shown to significantly improve LLMs’ performance on tasks requiring logical reasoning, planning, or multi-step decision making~\cite{kojima2022large, lyu2023faithful}. In the context of bug reproduction, CoT is particularly valuable as it enables LLMs to generate UI actions in an iterative, step-by-step manner rather than producing the entire sequence in a single pass. By reasoning over the current UI state, selecting the next action, and assessing its effect on the app, CoT ensures coherence across multi-step action generation. This structured reasoning process reduces the likelihood of incomplete or inconsistent action sequences, thereby enhancing the accuracy and reliability of LLM-driven bug reproduction.

\begin{figure}[t]
    \centering
    \includegraphics[width=\linewidth]{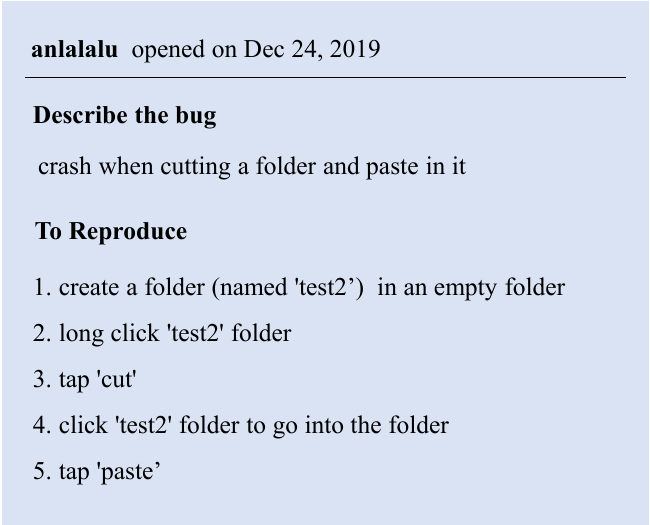}
    \caption{A real bug report (AmazeFileManager\#1796) used as the motivating example.}
    \label{fig:bug-report}
\end{figure}

\section{Motivation}

\begin{figure*}[t]
    \centering
    \includegraphics[width=\linewidth]{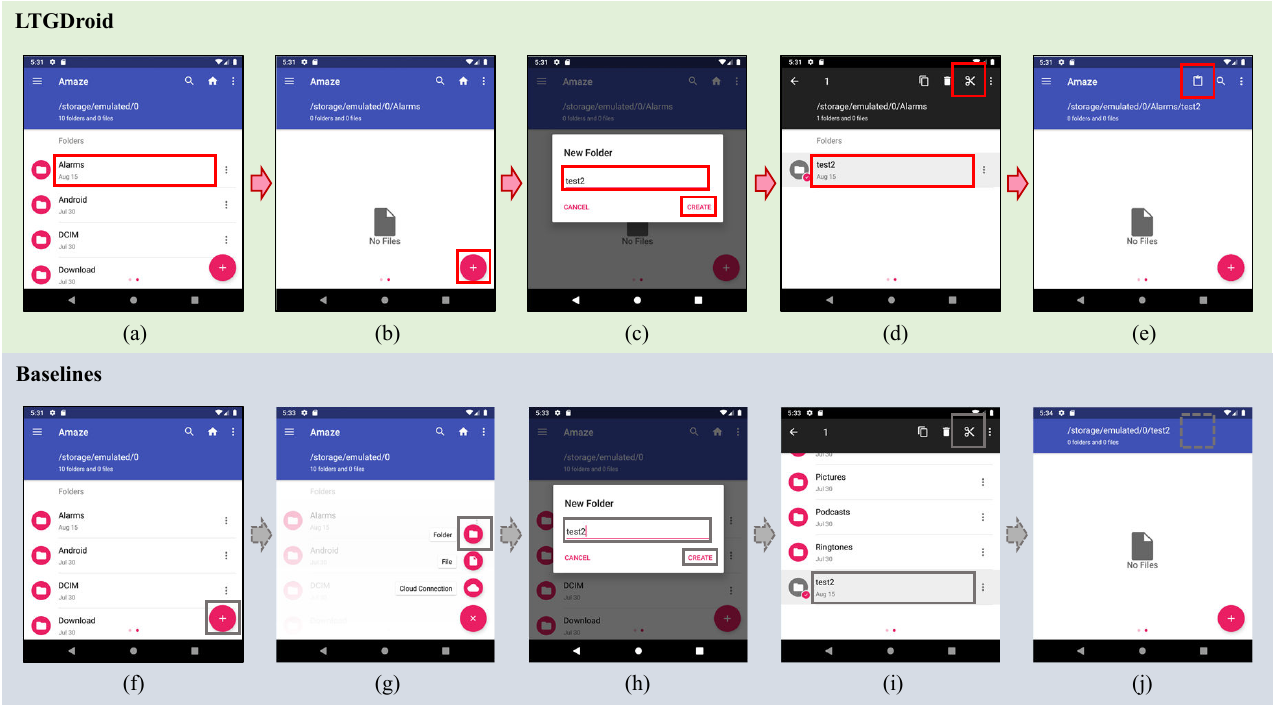}
    \caption{Reproduction actions generated by LTGDroid and baseline approaches for the motivating bug report.}
    \label{fig:reproduction-actions}
\end{figure*}

We use a real bug report shown in Fig.~\ref{fig:bug-report} to illustrate the limitations of existing methods and to motivate our work. The example is drawn from AmazeFileManager~\cite{AmazeFileManager}, a popular file management app on GitHub with over 5.7K stars. In Issue \textit{AmazeFileManager\#1796}~\footnote{\url{https://github.com/TeamAmaze/AmazeFileManager/issues/1796}}, the developer raised a bug report: when cutting a folder and pasting it into itself, the app crashes. Our goal is to transform the reproduction steps provided in the bug report (Fig.~\ref{fig:bug-report}) into the correct UI actions illustrated in Fig.~\ref{fig:reproduction-actions} (a)--(e). Specifically, (a) clicks the entry of the empty folder ``Alarms'', leading to (b), where the floating button at the bottom-right corner is clicked to open the creation menu and the option to create a new folder is selected. This brings up the dialog in (c), where the name ``test2'' is entered and the ``CREATE'' button is pressed, resulting in the newly created folder shown in (d). The folder is then long-pressed and cut using the top-right cut button. Finally, in (e), after entering the test2 folder, the paste button in the top-right corner is clicked, thereby completing all the UI actions required to reproduce the bug.

We apply the following two baseline approaches to the real bug report shown in Fig.~\ref{fig:bug-report}:

\begin{itemize}[leftmargin=*]
\item \textbf{AdbGPT~\cite{AdbGPT}}: The first approach that leverages the LLM for bug report reproduction. Specifically, AdbGPT identifies the S2R entities from the bug report and then uses these extracted entities as prompts, together with the available UI widgets, to determine the most suitable UI actions for replaying the S2R.

\item \textbf{ReBL~\cite{ReBL}}: The current state-of-the-art method for bug report reproduction. ReBL improves upon AdbGPT by eliminating the dependency on S2R extraction. It employs a feedback-driven mechanism that integrates the bug report, the current app state, and the history of reproduction actions to capture a richer UI context. This design enables ReBL to more effectively handle incomplete or ambiguous information in bug reports.
\end{itemize}

Although we provided both AdbGPT and ReBL with the complete S2R and all relevant information from the bug report, neither of them succeeded in reproducing the bug. The ``Baselines'' part in Fig.~\ref{fig:reproduction-actions} shows the UI actions generated by AdbGPT and ReBL when attempting to reproduce the step ``create a new folder in an empty directory.'' As shown, both methods overlooked the critical clue of the ``empty folder''. For example, when resolving ambiguous S2R, ReBL attempts to optimize AdbGPT by leveraging non-S2R information, including contextual details from the bug report itself, as well as the app’s current UI state and reproduction history. In the interface shown in Fig.~\ref{fig:reproduction-actions}(f), ReBL considers clicking the floating button at the bottom-right corner as the best option for creating a new folder, since this button might open a menu with such functionality. Furthermore, when exploring the interface in Fig.~\ref{fig:reproduction-actions}(j), ReBL tries to find a paste button to perform Step 5 from the bug report in Fig.~\ref{fig:bug-report}. However, since the paste button does not exist in this state, ReBL continues searching for alternative UI widgets, which causes the triggered actions to increasingly diverge from the intended reproduction path, eventually making the process unable to terminate in a timely manner.

To overcome these limitations, LTGDroid does not rely solely on the current UI state and reproduction history when selecting UI actions. Instead, LTGDroid enumerates all possible UI actions available on the current screen and executes them to observe their visual runtime behaviors. The extracted visual effects are then provided to the LLM, which uses them to select the action most likely to advance the bug reproduction task. 
For example, when LTGDroid reaches the interface shown in Fig.~\ref{fig:reproduction-actions}(a), it proactively attempts all possible UI actions on that screen (including clicking the empty folder ``Alarms'') to capture the corresponding visual runtime behaviors. Based on these observations, LTGDroid can determine that entering the empty folder first (Fig.~\ref{fig:reproduction-actions}(a)) and then clicking the floating button (Fig.~\ref{fig:reproduction-actions}(b)) is more likely to accomplish the intended step. LTGDroid then continues along this path until it reaches Fig.~\ref{fig:reproduction-actions}(e), where the UI action successfully triggers the bug.
Although this exploration incurs additional overhead to collect the visual effects of different UI actions, our experimental results demonstrate that such an investment significantly improves LTGDroid’s success rate in bug reproduction (see Section~\ref{sec:RQ2}).
\begin{figure*}[t]
    \centering
    \includegraphics[width=\textwidth]{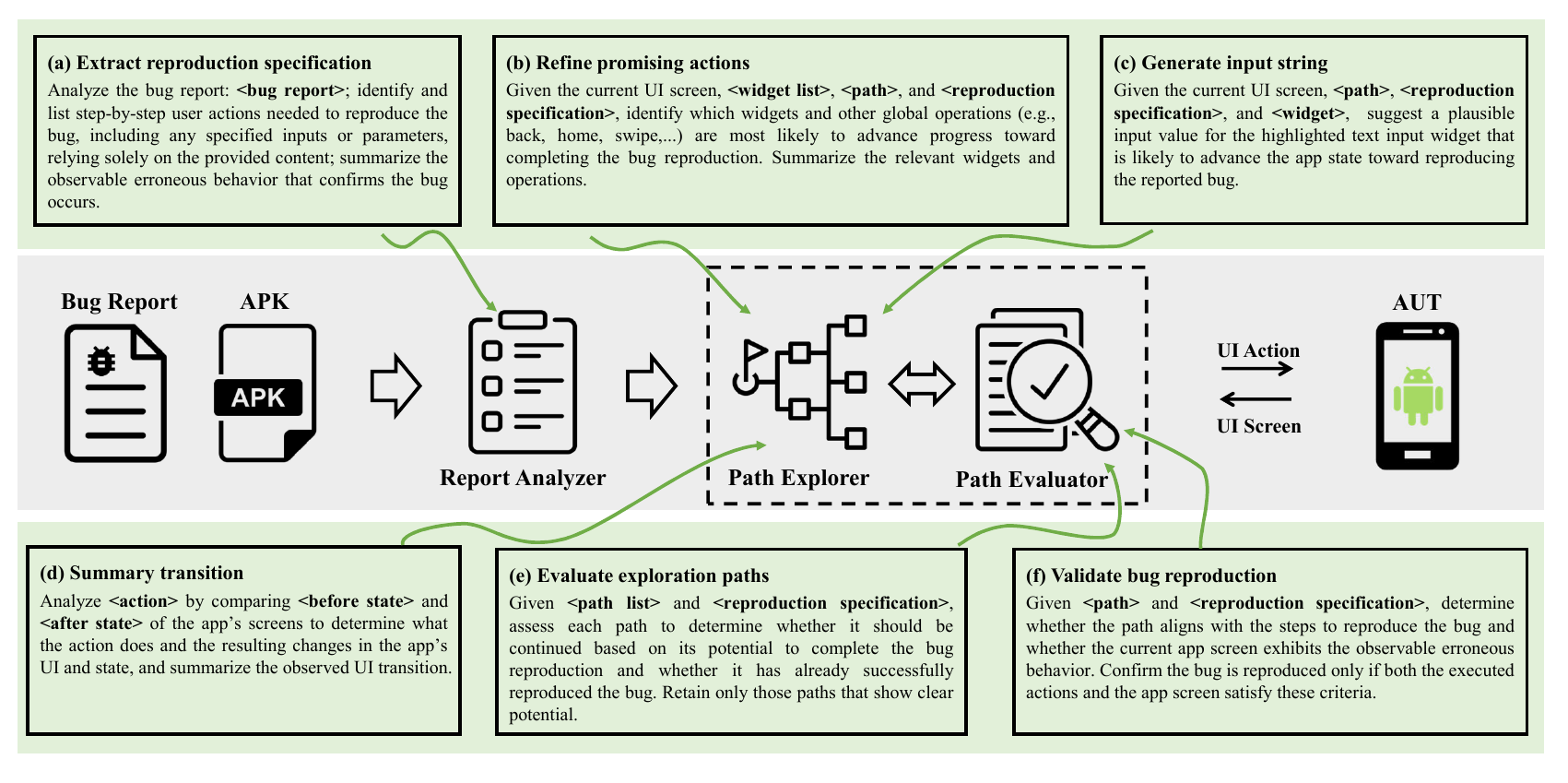}
    \caption{Overview of the workflow and prompt structures employed in LTGDroid for automated bug reproduction.}
    \label{fig:overview}
\end{figure*}

\section{Approach}

We propose LTGDroid, an automated bug reproduction technique for Android apps. Specifically, LTGDroid takes a bug report and the app under test (AUT) as input, and then automatically generates and executes a sequence of UI actions that trigger the bug described in the report. Essentially, the process of bug reproduction can be abstracted as starting from the initial state of the AUT and progressively exploring feasible paths of UI actions until reaching a bug whose behavior is consistent with the description in the bug report. We therefore model bug reproduction as a state-transition process that begins with an empty initial graph and dynamically expands during exploration.

Building on this idea, LTGDroid is designed with three core modules: Report Analyzer, Path Explorer, and Path Evaluator, as illustrated in Fig.~\ref{fig:overview}. First, the Report Analyzer parses the complete natural language bug report to extract the bug reproduction specification, including the steps to reproduce (S2R) and observable error symptoms, which serve as the goals for subsequent exploration and evaluation. Then, the Path Explorer and Path Evaluator iteratively generate and filter candidate UI actions based on the UI screenshots and the corresponding view hierarchy returned by the AUT, progressively exploring potential interaction paths until the Path Evaluator confirms that a path has successfully reproduced the bug.

In addition to the overall workflow, Fig.~\ref{fig:overview} also illustrates the structures of the prompts employed in each module. These prompt structures make use of a set of attributes (e.g., \textbf{\textless bug report\textgreater}, \textbf{\textless widget list\textgreater}) to represent placeholders within the prompts. The detailed descriptions of these attributes are summarized in Table~\ref{tab:prompt-attributes}. Due to space limitations, the figure and table only present the description of the prompts rather than their complete content. The full prompt details are publicly available in our GitHub repository~\cite{LTGDroid}.

\subsection{Report Analyzer}

Our goal is to achieve automated bug reproduction on a given AUT using the original bug report as input. However, the structure and content of bug reports are often highly variable: they may include S2R, observed failure symptoms, error logs, or even potential root-cause analyses. The completeness and level of detail largely depend on the reporter’s expertise. Relying solely on the S2R section may omit crucial information~\cite{ReBL}, while directly using the original bug report as the prompt introduces excessive noise, which not only interferes with LLMs' reasoning but also significantly increases token consumption.

To address this challenge, we adopt a compromise strategy. Instead of using only the S2R section or the original report, we leverage LLMs to analyze and interpret the complete bug report, distilling and complementing the S2R while also summarizing the observable error symptoms at the UI level. By leveraging the natural language understanding and reasoning capabilities of LLMs, the Report Analyzer preserves key information while filtering out redundant details, thus providing clearer and more reliable reproduction targets for the Path Explorer and Path Evaluator. Specifically, we design a structured prompting strategy (see Fig.~\ref{fig:overview}(a)) that enables the extraction of a precise exploration objective from the bug report.

\begin{table}[htbp]
	\centering
    \small
	\caption{Prompt Attribute Descriptions in LTGDroid}
	\label{tab:prompt-attributes}
	\begin{threeparttable}
		\begin{tabularx}{\linewidth}{p{1.7cm}|X}
			\toprule
			\textbf{Attribute}  & \textbf{Description} \\
			\midrule
            bug report & The original bug report, including the title, body, and comments in plain text. \\
			\midrule
            widget & The XML-formatted representation of a widget, derived from its view hierarchy properties to capture its semantic meaning in natural language.
            \textbf{Example:} \texttt{\textless EditText android:text="Enter Name"/\textgreater}. \\
            \midrule
            widget list & The list of widget elements in the current view hierarchy, each paired with its executable UI actions. \\
			\midrule
            path & The currently explored path, represented by all transitions along the path and reflecting the exploration history of the corresponding STG node; \newline \textbf{Example:} (1) Clicking ``Alarms'' opens the folder, updates the view to show its empty contents; (2) Clicking the floating action button expands a mini-menu with quick actions for creating a folder, file, or cloud connection. \\
			\midrule
            path list & The list of all ongoing exploration paths in the current iteration, where each element follows the same structure as \textless path\textgreater. \\
			\midrule
            reproduction specification & The S2R and observable error symptoms distilled by the Report Analyzer from the bug report, representing the exploration objective. \\
			\midrule
            action & The natural language description of a UI action;
            \textbf{Example:} input ``test2'' in widget \texttt{\textless EditText android:text="Enter Name"/\textgreater}. \\
			\midrule
            before state/ after state & The UI screen before/after executing an action, consisting of the screenshot and the name of the current activity. \\
			\bottomrule
		\end{tabularx}
	\end{threeparttable}
\end{table}

\subsection{Path Explorer} \label{sec:path-explorer}

In LTGDroid, we introduce a pre-exploration process that systematically collects and analyzes all possible UI actions and their corresponding state transitions based on the current UI state of the AUT. The goal of this process is not to directly complete bug reproduction but to generate a set of more comprehensive candidate exploration paths, which will later serve as input for the Path Evaluator. By performing a preliminary assessment of UI actions, LTGDroid provides a more targeted foundation for exploration, thereby enabling the Path Evaluator to retain only those paths most likely to trigger the target bug.

Specifically, the Path Explorer enumerates all executable UI actions from the current UI state and analyzes the effect of each action and the resulting visual changes in the app’s UI and state, which facilitates subsequent filtering in the Path Evaluator.
To enumerate all available UI actions, LTGDroid first parses the current XML-based view hierarchy to construct a widget tree and inspects each widget's attributes (e.g., \texttt{android:clickable}). Based on these attributes and widget types, it systematically derives a comprehensive set of feasible UI actions.
In total, LTGDroid supports six major categories of actions: tapping or long-pressing a UI widget (\texttt{Click}, \texttt{LongClick}); performing directional swipe gestures (\texttt{Swipe}); entering LLM-generated text (\texttt{InputText}; generated using the prompt shown in Fig.~\ref{fig:overview}(c)); switching between portrait and landscape orientation (\texttt{Rotate}); and invoking system-level key events (\texttt{Press}; e.g., \textit{Back}, \textit{Enter}, \textit{Delete}, \textit{Home}).
% In total, LTGDroid supports six major categories of actions, including tapping or long-pressing a UI element (\texttt{Click} and \texttt{LongClick}), performing directional swipe gestures (\texttt{Swipe}), entering text generated from a prompt (see Fig.~\ref{fig:overview}(c)) that integrates the current UI screen and exploration objective (\texttt{InputText}), switching the device orientation between portrait and landscape (\texttt{Rotate}), and invoking system-level key events (\texttt{Press}), such as \textit{Back}, \textit{Enter}, \textit{Delete}, and \textit{Home}.

% LTGDroid supports six major types of UI actions:

% \begin{itemize}
%   \item \textbf{Click}: Tap the center of a UI element.
%   \item \textbf{LongClick}: Press and hold the center of a UI element for one second.
%   \item \textbf{Swipe}: Perform swipe gestures in one of four directions (up, down, left, right).
%   \item \textbf{InputText}: Generate a suitable input string based on a prompt (see Fig.~\ref{fig:overview}(c)) that combines the UI screen and exploration objective and enter it into the target UI element.
%   \item \textbf{Rotate}: Switch the device orientation between landscape and portrait.
%   \item \textbf{Press}: Trigger a system key event, such as \textit{Back}, \textit{Enter}, \textit{Delete}, or \textit{Home}.
% \end{itemize}

A straightforward approach would be to evaluate every possible action in the current state. However, this method faces two major challenges in practice:
(1) A large portion of UI actions and their visual effects are irrelevant to the exploration objective, leading to inefficiency in path evaluation and potentially exceeding the LLM’s maximum token limit.
(2) The number of actions in a single UI state can be substantial, and evaluating them one by one would incur prohibitive time costs and token consumption.

To address these issues, LTGDroid invokes the LLM to analyze the semantic information of the current screen and filter out actions unrelated to the exploration objective. This reduces the input size for the action planner while preserving relevance. In practice, we design a structured prompt (see Fig.~\ref{fig:overview}(b)) that combines the UI screen, the exploration objective (i.e., the \textless reproduction specification\textgreater), and the exploration path context to refine the action set. The process is detailed in Algorithm~\ref{alg:pre-exploration}. In line 2, LTGDroid first extracts all executable actions from the current UI state (denoted as $actions$). LTGDroid then invokes the LLM to discard actions that are unlikely to contribute to the objective (lines 3).

\begin{algorithm}[t]
    \caption{Pre-Exploration of UI Actions}
    \label{alg:pre-exploration}

    \KwIn{$state$: Current UI state of the AUT, $objective$: Exploration Objective}
    \KwOut{$transitions$: Set of candidate UI transitions}
    \SetAlgoLined

    $transitions \gets \emptyset$\;
    $actions \gets \text{extractActions}(state)$\;
    $actions \gets \text{filterActions}(actions, objective)$\;
    
    \ForEach{$action \in actions$}{
        $screen\_before \gets \text{captureScreen}()$\;
        \text{execute}($action$)\;
        $screen\_after \gets \text{captureScreen}()$\;
        $transition \gets \text{summTrans}(action, screen\_before, screen\_after)$\;
        $transitions \gets transitions \cup \{transition\}$\;
        \text{rollback}($state$)\;
    }
    
    \Return $transitions$\;
\end{algorithm}

After refining the action set, LTGDroid executes each remaining action and captures the UI screens before and after execution. These, together with contextual information of the action, are summarized as UI transitions and added to the $transitions$ collection for subsequent use by the Path Evaluator (lines 4--11).
% To maintain independence among actions, LTGDroid rolls back to the original snapshot after each execution (line 12).
To maintain independence among actions, LTGDroid rolls back the emulator to a previously saved snapshot after each execution (line 10).
For example, in the bug reproduction task shown in Fig.~\ref{fig:bug-report}, LTGDroid first extracted all executable actions in the current UI state (Fig.~\ref{fig:reproduction-actions}(a)). LTGDroid then employed the LLM to filter out irrelevant ones. The retained candidates included actions such as ``clicking the floating button in the bottom-right corner'' and ``clicking the entry for the Alarm folder''. LTGDroid then executed these actions iteratively, recorded the corresponding UI transitions, and added them to the interaction history of the corresponding exploration path for subsequent evaluation.

In practice, we leverage an LLM to summarize the UI transitions by designing a structured prompt (see Fig.~\ref{fig:overview}(d)). 
Importantly, we focus on observable UI changes rather than system-level metrics (e.g., memory usage or CPU load), as our goal is to capture semantically meaningful state evolutions from the user’s perspective. To this end, we instruct the LLM to analyze the captured UI screenshots together with the action-related contextual information and generate a structured summary at multiple levels of granularity. Specifically, the LLM describes both what the action does and the resulting changes in the app’s UI and state, with emphasis on phenomena such as widget additions or removals, visibility updates, content modifications, layout shifts, and screen or Activity switches. This structured characterization ensures that the extracted UI transitions are precise, consistent, and directly aligned with the needs of the subsequent path evaluation process.

Notably, although the LLM may occasionally filter out actions that are actually relevant to the exploration objective, our experiments indicate that such cases only rarely affect LTGDroid's overall effectiveness (see Reason \#1 in Section \ref{sec:RQ1}).

\subsection{Path Evaluator}

Based on the exploration results provided by the Path Explorer, the Path Evaluator filters the ongoing exploration paths, pruning branches that deviate from the exploration objective and retaining only the most promising ones for further expansion in the next iteration. This yields a BFS-like bounded exploration process that continues until the evaluator determines that a path has successfully fulfilled the exploration objective.
For illustration, consider the real bug reproduction task illustrated in Fig.~\ref{fig:bug-report}. LTGDroid first obtains UI transitions relevant to the exploration objective through pre-assessment (see Section~\ref{sec:path-explorer}) in the current UI state (Fig.~\ref{fig:reproduction-actions}(a)). The evaluator then analyzes these paths: the path leading from (a) to (b), entering an empty folder, is considered beneficial for reaching the objective and thus retained for further iteration, whereas the path from (a) directly clicking the floating button to (g), entering a menu, is deemed irrelevant and is pruned. This iterative exploration and evaluation continues until the exploration path from (a) to (e) is confirmed to reproduce the bug, at which point the process ends.

Specifically, we design a structured prompt (see Fig.~\ref{fig:overview}(e)) that combines the exploration objective (i.e., the \textless reproduction specification\textgreater) with the ongoing exploration paths to assess whether a path is worth further exploration and whether it has already triggered the target bug. The evaluator then retains the most promising paths and forwards them to the Path Explorer for the next round of expansion.
% Moreover, considering that the LLM may have an incomplete understanding of app-specific behaviors, we enrich the prompt with auxiliary knowledge to guide its judgment.
% For instance, the prompt instructs the evaluator to account for necessary initialization steps before the bug can be reproduced, such as granting permissions, dismissing update dialogs, progressing through or skipping onboarding screens, or consenting to data collection prompts. Similarly, to prevent the exploration from wasting resources on unproductive paths, the prompt explicitly directs the evaluator to identify and discard paths that fall into repetitive loops, as such paths are unlikely to contribute toward reaching the reproduction objective. By embedding these instructions, the evaluator is guided to retain paths that are both relevant and practical for continued exploration.

To mitigate the risk of the LLM prematurely determining that an exploration path has reached the objective due to inherent randomness or limited contextual information, we incorporate an additional verification step before concluding the process. Specifically, upon the evaluator’s initial identification of a path as satisfying the exploration objective, we validate this outcome by employing a dedicated prompt (see Fig.~\ref{fig:overview}(f)) to examine the corresponding UI screen and ensure that the observed behavior aligns with the expected bug manifestation.
For crash-related bugs, the evaluator monitors error logs to confirm the occurrence of the crash.
This additional verification reduces the chance of inaccurate conclusions caused by LLM limitations and helps ensure that identified paths correspond authentically to the target bug reproduction.

\section{Evaluation}

To evaluate LTGDroid, we formulate three key research questions:

\begin{itemize}[leftmargin=*]
    \item \textbf{RQ1:} How effective and efficient is LTGDroid in reproducing bug reports?
    \item \textbf{RQ2:} How do the modules of LTGDroid affect its effectiveness and efficiency?
    \item \textbf{RQ3:} How does LTGDroid perform compared to three baseline methods in terms of effectiveness and efficiency?
\end{itemize}

\subsection{Dataset}

\begin{table}[htbp]
    \centering
    \tiny
    \caption{App Subjects Used in Our Evaluation (K=1,000, M=1,000,000)}
    \label{tab:app-subjects}
    \begin{threeparttable}
        \begin{tabular}{l|l|c|c|c}
            \toprule
            \textbf{App} & \textbf{Feature Description} & \makecell{\textbf{\#Google Play}\\\textbf{Installations}} & \makecell{\textbf{\#Github}\\\textbf{Stars}} & \makecell{\textbf{\#Lines of}\\\textbf{Code}}\\
            \midrule
            ActivityDiary & Activity Scheduler & 1K+ & 75 & 18.2K \\
            AmazeFileManager & Lightweight File Manager & 1M+ & 5.7K & 114.8K \\
            AndrOBD & OBD Car Diagnostics & 1K+ & 1.7K & 61.9K \\
            android-noad-music-player & Local Music Player & 10K+ & 69 & 13.4K \\
            android-obd-reader & OBD Car Diagnostics & - & 827 & 28.1K \\
            anglers-log & Fishing Log \& Analytics & 100K+ & 31 & 134.2K \\
            Anki-Android & Spaced Repetition Flashcards & 10M+ & 9.9K & 282K \\
            AntennaPod & Podcast Manager & 1M+ & 7.2K & 107.8K \\
            AnyMemo & Spaced Repetition Flashcards & 100K+ & 153 & 33.4K \\
            APhotoManager & Local Photo Gallery & - & 231 & 48.2K \\
            AsciiCam & Real-Time ASCII Camera & 100K+ & 134 & 3.3K \\
            Birthdroid & Birthday Reminder & - & 14 & 2.9K \\
            FastAdapter & RecyclerView Adapter & - & 3.9K & 15K \\
            fastnfitness & Fitness Tracker & 5K+ & 302 & 36.2K \\
            FirefoxLite & Web Browser & 5M+ & 292 & 146.6K \\
            focus-android & Web Browser & 10M+ & 2.1K & 81.2K \\
            gnucash-android & Personal Finance Manager & 10K+ & 1.2K & 53.3K \\
            gpstest & GPS Test \& Diagnostics & 1M+ & 2K & 26.8K \\
            jtxBoard & Journal, Notes Manager & 10K+ & 499 & 296K \\
            KISS & Minimalist Launcher & 100K+ & 3.2K & 44.2K \\
            kiwix-android & Offline Wikipedia Reader & 1M+ & 1.1K & 80.6K \\
            LibreNews-Android & News Aggregator & - & 36 & 2.5K \\
            LocalMaterialNotes & Local Markdown Notes & 1K+ & 239 & 21.5K \\
            markor & Markdown \& To-Do Notes & - & 4.6K & 65.7K \\
            MaterialFBook & Facebook Client & - & 139 & 9.4K \\
            MaterialFiles & File Manager & 1M+ & 7.3K & 83.7K \\
            Memento-Calendar & Calendar \& Event Manager & 10K+ & 210 & 31K \\
            mhabit & Habit Tracker \& Goal Setter & - & 767 & 76.4K \\
            microMathematics & Scientific Calculator & 50K+ & 612 & 80.6K \\
            NewPipe & Lightweight YouTube Client & - & 34.8K & 169.3K \\
            NewsBlur & RSS Reader & 50K+ & 7.1K & 418.3K \\
            NotallyX & Minimalistic Note Manager & 1K+ & 334 & 309.6K \\
            Omni-Notes & Note Manager & 100K+ & 2.8K & 46.4K \\
            OpenCalc & Scientific Calculator & 100K+ & 1.3K & 16K \\
            osmdroid & OpenStreetMap Tools & 10K+ & 3K & 132.4K \\
            privacy-friendly-todo-list & To-Do List & - & 110 & 16.4K \\
            privacy-friendly-weather & Weather & - & 121 & 226.9K \\
            qksms & Customizable SMS Manager & 1M+ & 4.6K & 59.5K \\
            screenrecorder & Screen Recording & 50K+ & 122 & 7.2K \\
            thunderbird-android & Email Client & 5M+ & 12.4K & 314.1K \\
            Transportr & Public Transport Tracker & 10K+ & 1.1K & 40.5K \\
            trickytripper & Trip Planner & 5K+ & 47 & 3.5M \\
            uhabits & Habit Tracker \& Goal Setter & 5M+ & 9K & 53.3K \\
            URLCheck & Batch URL Checker & 100K+ & 1.5K & 23.7K \\
            WiFiAnalyzer & Wi-Fi Analyzer & 10M+ & 4.1K & 25K \\
            \bottomrule
        \end{tabular}
    \end{threeparttable}
\end{table}

In this study, we constructed a dataset for Android bug reproduction. Specifically, we first integrated the evaluation datasets used in four existing studies: ReCDroid~\cite{ReCDroid}, ReproBot~\cite{ReproBot}, AdbGPT~\cite{AdbGPT}, and ReBL~\cite{ReBL}, together with two manually reproduced bug report datasets, AndroR2~\cite{Andror2} and Themis~\cite{Themis}.
During the data cleaning process, we removed duplicate entries and reports not hosted on GitHub (since our bug reports were collected via the GitHub REST API~\cite{GithubAPI}), and further filtered out unusable samples through manual reproduction. Unusable samples included reports with broken links, missing corresponding APK files, dependencies on special initialization resources (e.g., account login, importing audio/video or SMS records). After filtering, 48 valid samples remained, including 37 crash reports and 11 non-crash reports. Since many reports in the above datasets corresponded to older defects, we additionally collected 27 new bug reports submitted in 2024 or later from some of the apps in the existing datasets as well as several popular GitHub open-source apps (Stars \textgreater\ 200). These included 14 crash reports and 13 non-crash reports.

We constructed a dataset of 75 bug reports from 45 different apps (see Table~\ref{tab:app-subjects}), including 51 crash reports and 24 non-crash reports. For each report, we manually annotated the number of ground-truth UI actions required for successful reproduction. Detailed information for all bug reports is provided in Table~\ref{tab:bug-reports}.
The distribution of these actions is broad, indicating no bias toward overly simple or complex bugs: 28 reports required 1–5 actions, 34 reports 6–10 actions, 11 reports 11–15 actions, and 2 reports more than 15 actions.
Notably, all bug reports in our dataset are kept in plain-text form to ensure compatibility with existing evaluation baselines. If needed, LTGDroid can be easily extended to support image-based bug reports.

\begin{table}[htbp]
    \centering
    \tiny
    \caption{Bug Reports Used in Our Evaluation}
    \label{tab:bug-reports}
    \begin{threeparttable}
        % \begin{tabular}{c|p{3.5cm}p{1.3cm}c||c|p{3.5cm}p{1.3cm}c}
        \begin{tabularx}{0.9\linewidth}{>{\centering\arraybackslash}p{0.8cm}|X|p{1.5cm}|>{\centering\arraybackslash}p{0.8cm}}
            \toprule
            \textbf{ID} & \textbf{Report} & \textbf{Type} & \textbf{\#GA}\\
            \midrule
1 & ActivityDiary\#118 & Crash & 15 \\
2 & ActivityDiary\#285 & Crash & 7 \\
3 & AmazeFileManager\#1796 & Crash & 10 \\
4 & AmazeFileManager\#4185 & Crash & 4 \\
5 & AndrOBD\#144 & Crash & 6 \\
6 & android-noad-music-player\#1 & Crash & 2 \\
7 & android-obd-reader\#22 & Crash & 7 \\
8 & Anki-Android\#10584 & Crash & 10 \\
9 & Anki-Android\#4586 & Crash & 6 \\
10 & Anki-Android\#4707 & Crash & 6 \\
11 & Anki-Android\#4977 & Crash & 3 \\
12 & Anki-Android\#5638 & Crash & 4 \\
13 & Anki-Android\#6432 & Crash & 32 \\
14 & AnyMemo\#422 & Crash & 4 \\
15 & AnyMemo\#440 & Crash & 7 \\
16 & APhotoManager\#116 & Crash & 2 \\
17 & AsciiCam\#17 & Crash & 3 \\
18 & Birthdroid\#13 & Crash & 6 \\
19 & FastAdapter\#394 & Crash & 1 \\
20 & fastnfitness\#142 & Crash & 9 \\
21 & fastnfitness\#293 & Crash & 15 \\
22 & FirefoxLite\#5085 & Crash & 5 \\
23 & gnucash-android\#664 & Crash & 15 \\
24 & jtxBoard\#1727 & Crash & 5 \\
25 & kiwix-android\#990 & Crash & 7 \\
26 & LibreNews-Android\#22 & Crash & 6 \\
27 & LibreNews-Android\#23 & Crash & 7 \\
28 & LocalMaterialNotes\#385 & Crash & 9 \\
29 & markor\#194 & Crash & 8 \\
30 & MaterialFBook\#224 & Crash & 1 \\
31 & Memento-Calendar\#169 & Crash & 9 \\
32 & microMathematics\#39 & Crash & 4 \\
33 & NewPipe\#11914 & Crash & 8 \\
34 & NewPipe\#11915 & Crash & 10 \\
35 & NewsBlur\#1053 & Crash & 4 \\
36 & NotallyX\#517 & Crash & 6 \\
37 & NotallyX\#569 & Crash & 3 \\
38 & NotallyX\#609 & Crash & 4 \\
39 & Omni-Notes\#377 & Crash & 2 \\
40 & Omni-Notes\#745 & Crash & 10 \\
41 & OpenCalc\#493 & Crash & 2 \\
42 & OpenCalc\#532 & Crash & 2 \\
43 & osmdroid\#1030 & Crash & 9 \\
44 & privacy-friendly-weather\#61 & Crash & 4 \\
45 & qksms\#585 & Crash & 6 \\
46 & screenrecorder\#25 & Crash & 8 \\
47 & thunderbird-android\#3255 & Crash & 4 \\
48 & Transportr\#972 & Crash & 5 \\
49 & trickytripper\#42 & Crash & 11 \\
50 & uhabits\#1545 & Crash & 10 \\
51 & URLCheck\#493 & Crash & 11 \\
52 & anglers-log\#151 & Non-crash & 11 \\
53 & anglers-log\#347 & Non-crash & 8 \\
54 & Anki-Android\#5753 & Non-crash & 7 \\
55 & AntennaPod\#7611 & Non-crash & 3 \\
56 & AntennaPod\#7866 & Non-crash & 5 \\
57 & focus-android\#3297 & Non-crash & 13 \\
58 & gpstest\#404 & Non-crash & 12 \\
59 & KISS\#1481 & Non-crash & 11 \\
60 & LocalMaterialNotes\#327 & Non-crash & 7 \\
61 & LocalMaterialNotes\#415 & Non-crash & 19 \\
62 & LocalMaterialNotes\#416 & Non-crash & 4 \\
63 & markor\#1020 & Non-crash & 15 \\
64 & markor\#331 & Non-crash & 8 \\
65 & MaterialFiles\#1281 & Non-crash & 8 \\
66 & MaterialFiles\#1392 & Non-crash & 5 \\
67 & Memento-Calendar\#7 & Non-crash & 5 \\
68 & mhabit\#270 & Non-crash & 9 \\
69 & OpenCalc\#448 & Non-crash & 6 \\
70 & OpenCalc\#510 & Non-crash & 8 \\
71 & OpenCalc\#524 & Non-crash & 5 \\
72 & privacy-friendly-todo-list\#158 & Non-crash & 11 \\
73 & thunderbird-android\#3971 & Non-crash & 6 \\
74 & uhabits\#2112 & Non-crash & 9 \\
75 & WiFiAnalyzer\#222 & Non-crash & 4 \\
            \bottomrule
        \end{tabularx}
    \end{threeparttable}
\end{table}

\subsection{Experimental Setup}

Our experiments were conducted on a Mac mini equipped with an Apple M4 chip (10-core CPU) and 32 GB of memory. The experimental environment adopted the official Android emulator configured with an Android 9.0 ARM system image, 4 CPU cores, 4 GB of memory, and hardware acceleration enabled.  

It is worth noting that, throughout the evaluation, we aimed to faithfully replicate the real-world scenarios faced by developers when reproducing bug reports. To this end, the experiments only automated the installation and launch of applications without relying on any manually crafted initialization scripts. Furthermore, to ensure fairness, all components involving LLMs uniformly employed GPT-4.1, a multimodal model developed by OpenAI.

\subsubsection{Evaluation Metrics}

We evaluated all tools from two perspectives: \emph{effectiveness} and \emph{efficiency}.

Effectiveness was assessed by the success rate (\textbf{SR}) formulated as follows:
\begin{equation}
\text{SR} = \frac{\text{\# of Successfully Reproduced Bug Reports}}{\text{\# of Bug Reports in Dataset}}
\end{equation}

Efficiency was assessed using four metrics: the average number of UI actions (\textbf{Actions}), execution time (\textbf{Time}), LLM token consumption (\textbf{Tokens}), and the corresponding monetary cost (\textbf{Cost}).
We took an upper bound of 100 UI actions and a maximum execution time of 60 minutes. Any attempt that exceeded either threshold was treated as a failure. Notably, the most complex bug in our dataset requires only 32 UI actions to reproduce manually, which is far below our predefined limit.

To ensure the reliability of our evaluation, three authors of this paper, each with at least two years of Android development experience, independently reviewed the execution results to determine whether the crashes or non-crash errors described in the bug reports were successfully reproduced. Each execution result was independently examined by two authors, and when their assessments differed, the third author re-evaluated the case to resolve the inconsistency.
To further mitigate randomness, each experiment was executed three times, and we report the average results across all runs for all metrics.

\subsubsection{Setup for RQ1}

To evaluate the effectiveness and efficiency of LTGDroid, we conducted experiments on the dataset of 75 real-world bug reports. We further analyzed all failure cases to understand the underlying causes and identify potential areas for improvement.

\subsubsection{Setup for RQ2}

To assess the contribution of each module in LTGDroid, we conducted an ablation study by comparing the full system with four variants, each disabling one core module. \textbf{LTGDroid w/o RA} removes the bug report analyzer, preventing the system from extracting structured reproduction specifications from the original bug report. \textbf{LTGDroid w/o AE} disables UI action enumeration, preventing the system from pre-assessing the effects of multiple UI actions on each screen. \textbf{LTGDroid w/o TA} excludes UI transition assessment, preventing the system from analyzing and recording the effects of UI actions. \textbf{LTGDroid w/o VB} removes bug behavior verification, preventing the system from confirming whether a candidate execution has triggered the target bug. By comparing these variants against the full LTGDroid system, we quantify how each module contributes to the effectiveness and efficiency.

\subsubsection{Setup for RQ3}

We selected three state-of-the-art LLM-based automated bug reproduction methods as baselines: \textbf{AdbGPT}~\cite{AdbGPT}, \textbf{ReBL}~\cite{ReBL}, and a visual variant of ReBL (\textbf{ReBL-visual}) that incorporates UI screenshots as additional visual context. Notably, since ReBL is already an improvement over AdbGPT, we did not construct a separate AdbGPT-visual variant. Following the same experimental setup as in previous studies, we set an upper limit of 100 UI actions for all methods to ensure a fair performance evaluation, and analyzed their reproduction effectiveness and efficiency within a reasonable action budget. All results were manually verified to ensure the reliability of the comparative analysis.  

It is worth noting that methods such as Yakusu~\cite{Yakusu}, ReCDroid~\cite{ReCDroid}, ReproBot~\cite{ReproBot}, ScopeDroid~\cite{ScopeDroid}, and Roam~\cite{Roam}, which are not LLM-based, rely on extracting S2R entities from bug reports and matching them to corresponding UI actions. However, most bug reports do not document the initialization steps required when launching a freshly installed app (e.g., permission grants, onboarding screens). Therefore, these methods typically require manually crafted initialization scripts to operate in practice. Moreover, they generally focus only on crash-related bugs and lack support for non-crash bugs. For these reasons, we did not include them as direct comparison baselines in this study.

\subsection{Results for RQ1} \label{sec:RQ1}

As shown in Table~\ref{tab:evaluation-results}, in terms of effectiveness, LTGDroid achieves a bug reproduction success rate of 88.82\% on crash cases and 84.71\% on non-crash cases, resulting in an overall success rate of 87.51\%.
In terms of efficiency, for each successfully reproduced bug, LTGDroid required an average of 27.48 UI actions, 67.07K tokens, \$0.27, and 20.45 minutes to complete the reproduction process.
These results indicate that LTGDroid is both robust and generalizable, achieving a high bug reproduction success rate with reasonable cost.
We further analyzed the 10 reproduced bug reports that LTGDroid failed to reproduce, including 6 crash reports and 4 non-crash reports. We identified the following four main causes.

\begin{figure}[t]
    \centering
    \includegraphics[width=\linewidth]{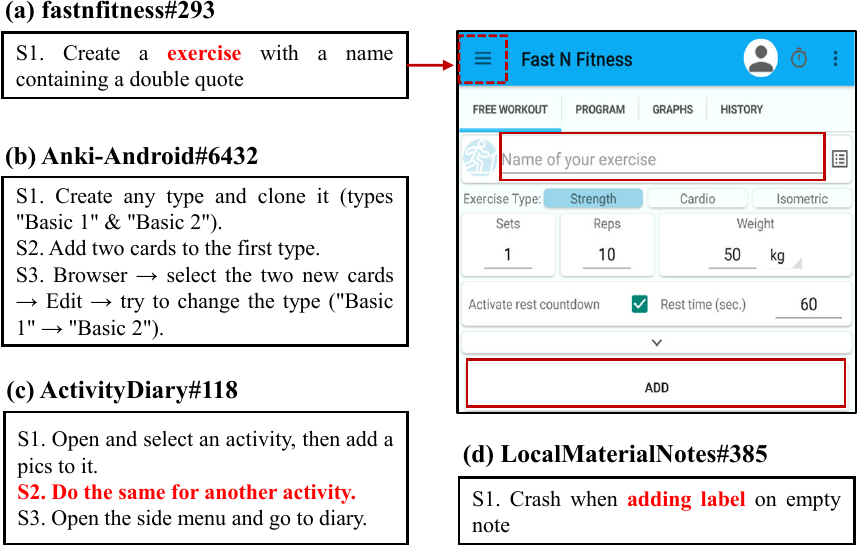}
    \caption{Representative examples of LTGDroid's bug reproduction failures.}
    \label{fig:failures}
\end{figure}

\noindent \textbf{Reason \#1: Incorrectly filtering out relevant UI actions.} LTGDroid relies on the LLM to filter out actions deemed irrelevant to the exploration goal, avoiding excessive attempts during the pre-assessment phase. However, due to the LLM's limited understanding of the app and bug report, it may mistakenly exclude critical actions. For example, in \textit{fastnfitness\#293} (Fig.~\ref{fig:failures}(a)), the LLM excluded the correct UI action, which involved clicking the top-left menu button leading to the exercise page, while retaining actions that appeared reasonable but were actually incorrect, such as entering and creating exercise records. We found only 1 such case in our evaluation.

\noindent \textbf{Reason \#2: Insufficient awareness of reproduction progress.} LTGDroid sometimes struggles to accurately track the status of each reproduction step. First, when a reproduction involves many steps, the LLM may misalign the historical context with the reproduction steps, leading to confusion about progress and repeated execution of completed steps. For instance, in \textit{Anki-Android\#6432} (Fig.~\ref{fig:failures}(b)), the LLM misinterpreted previously completed steps due to the large number of steps in the S2R, causing repeated actions. Second, when step descriptions are unclear, the LLM may have difficulty determining whether a step has been fully completed. This issue is illustrated by \textit{ActivityDiary\#118} (Fig.~\ref{fig:failures}(c)), where S2 required repeating the S1 actions on another activity, but the LLM prematurely proceeded to S3 before fully completing S2. We found 6 such cases in our evaluation.

\noindent \textbf{Reason \#3: Inability to infer and complete missing initialization steps.} When critical initialization steps are absent from the bug report, LTGDroid may fail to reproduce the bug. For example, in \textit{LocalMaterialNotes\#385} (Fig.~\ref{fig:failures}(d)), the bug report did not mention the need to create at least one label, so LTGDroid did not realize it had to perform this step, resulting in reproduction failure. We found 1 such case in our evaluation.

\noindent \textbf{Reason \#4: Limitations of the testing framework.} We found 2 bugs related to special operations that are not yet supported by LTGDroid. For instance, \textit{Memento-Calendar\#169} requires a swipe within a specific coordinate range, which LTGDroid cannot perform. Similarly, \textit{trickytripper\#42} requires a long-click on a UI element marked as non-longclickable, but LTGDroid strictly follows the view hierarchy properties and thus cannot execute this action.

Overall, the reproduction failures primarily stem from the LLM's limitations in understanding and reasoning, as well as the testing framework's operational constraints. These challenges highlight difficulties in handling complex or incomplete bug reports and suggest directions for future improvements in reasoning capability and test framework support.

\begin{mybox}
\textbf{Answer of RQ1:} LTGDroid achieves an overall success rate of 87.51\%. For each successfully reproduced bug, it requires an average of 27.48 UI actions, 67.07K tokens, \$0.27, and 20.45 minutes. Overall, the results demonstrate that LTGDroid is robust and generalizable while maintaining reasonable cost.
\end{mybox}

\subsection{Results for RQ2} \label{sec:RQ2}

% Table~\ref{tab:evaluation-results} presents the experimental results of LTGDroid without the pre-assessment mechanism (\textbf{LTGDroid w/o Pre}). In terms of effectiveness, \textbf{LTGDroid w/o Pre} successfully reproduced only 51 out of 75 bug reports, including 38 of 51 crash reports (74.51\%) and 13 of 24 non-crash functional reports (54.17\%), resulting in an overall success rate of 68.00\%, which represents a 21.54\% decrease compared to the full version of LTGDroid. This indicates that without pre-assessment, the system cannot verify the actual effects of UI actions in advance and must rely solely on the predictions of the LLM, significantly reducing the reproduction success rate. 

% In terms of efficiency, \textbf{LTGDroid w/o Pre} required an average of 9.20 UI actions to complete a single bug reproduction, which is 17.90 fewer actions than the full version. This is because in each iteration, the ablated version directly selects and executes a single UI action without pre-executing multiple candidate actions, thereby reducing the number of actions needed to complete bug reproduction. These results indicate that providing information about visual changes, despite incurring additional overhead in terms of time and tokens, is still effective in improving the success rate of bug reproduction tasks.

Table~\ref{tab:evaluation-results} presents the comparison among LTGDroid and its variants.
First, removing any critical component leads to a noticeable decrease in the reproduction success rate, among which disabling the UI transition assessment module (\textbf{LTGDroid w/o TA}) causes the largest drop. Its overall SR is only 51.11\%, as the absence of the module prevents the system from predicting the consequences of UI actions during reproduction. This variant also consumes the fewest tokens, mainly because it can only reproduce relatively simple bugs.

Second, when the bug report analyzer is removed (\textbf{LTGDroid w/o RA}), the system struggles to reliably extract structured reproduction specifications from raw bug reports, resulting in an overall SR of 62.67\%. Both token consumption and execution time of this variant increase significantly because unstructured and noisy reports take up a large amount of the available context budget, thereby reducing the efficiency of action planning.

Third, removing the bug behavior verification module (\textbf{LTGDroid w/o BV}) decreases the overall SR to 63.07\%, as the system can no longer reliably verify whether the target bug has been triggered, leading to incorrect judgments and premature termination of the reproduction process.

Fourth, removing the UI action enumeration module (\textbf{LTGDroid w/o AE}) results in an overall success rate of 68.44\%, indicating that the absence of UI action enumeration reduces the effectiveness of UI action evaluation during the reproduction process.

Notably, this variant achieves substantially lower execution time, as disabling action enumeration avoids the pre-assessment of multiple UI actions, which relies on emulator snapshots for state restoration and is time-consuming.
In summary, each functional module in LTGDroid is essential not only for ensuring a high success rate but also for balancing efficiency factors, together forming an optimal automated bug reproduction process.

\begin{mybox}
	\textbf{Answer of RQ2:} Each functional module in LTGDroid is essential for effective and efficient automated bug reproduction, as removing any component degrades success rate or efficiency. Notably, the UI transition assessment module is the most critical, guiding the interaction flow toward successful reproduction.
\end{mybox}

\begin{table*}[t]
	\centering
	\caption{Evaluation results of LTGDroid and baselines.}
    \small
	\label{tab:evaluation-results}
	\begin{threeparttable}
		\begin{tabular}{|l|c|c|c|c|c|c|c|c|}
			\toprule
			\textbf{Method} &\textbf{Crash SR (\%)} & \textbf{Non-crash SR (\%)} & \textbf{Overall SR (\%)} & \textbf{Actions (\#)} & \textbf{Tokens (K)} & \textbf{Cost (\$)} & \textbf{Time (m)} \\
			\midrule
			AdbGPT & 11.76 & 16.67 & 13.33 & 4.10 & 25.16 & 0.05 & 1.40 \\
			ReBL & 67.98 & 38.88 & 58.67 & 10.98 & 46.36 & 0.10 & 9.13 \\
			ReBL-visual & 70.59 & 52.79 & 64.89 & 9.39 & 53.95 & 0.11 & 10.09 \\
            \midrule
            LTGDroid w/o RA & 71.90 & 43.04 & 62.67 & 21.04 & 65.80 & 0.25 & 21.84\\
			LTGDroid w/o AE & 75.16 & 54.17 & 68.44 & 9.18 & 46.16 & 0.18 & 7.96 \\
            LTGDroid w/o TA & 58.82 & 34.71 & 51.11 & 25.74 & 31.31 & 0.14 & 18.40 \\
            LTGDroid w/o BV & 71.18 & 45.83 & 63.07 & 20.13 & 56.89 & 0.23 & 18.62 \\
			\midrule
			LTGDroid & 88.82 & 84.71 & 87.51 & 27.48 & 67.07 & 0.27 & 20.45 \\
			\bottomrule
		\end{tabular}
	\end{threeparttable}
\end{table*}

\subsection{Results for RQ3}

Table~\ref{tab:evaluation-results} presents the comparison results between LTGDroid and the baseline methods. Evaluated on 75 bug reports, including 51 crash reports and 24 non-crash reports, LTGDroid outperformed state-of-the-art LLM-based methods (\textbf{AdbGPT}~\cite{AdbGPT}, \textbf{ReBL}~\cite{ReBL}, and \textbf{ReBL-visual}) in terms of effectiveness. Its overall reproduction success rates increased by 556.30\%, 49.16\%, and 34.85\% respectively, and achieved the best performance for both crash and non-crash bug reports.

Notably, the bugs successfully reproduced by LTGDroid tend to involve higher interaction complexity, as reflected by a greater number of ground-truth UI actions required for reproduction (6.40 on average), compared to AdbGPT (3.50), ReBL (5.77), and ReBL-visual (6.29). 
The successful reproduction traces reported by LTGDroid contain an average of 6.89 UI actions per bug, which is close to the ground-truth average of 6.40, indicating that LTGDroid can report realistic and concise reproduction paths.
For the other baselines, the length of a reproduction trace directly corresponds to the number of executed UI actions. As a result, the reproduction traces they produce exhibit much larger deviations from the ground-truth action counts, which incurs additional manual verification effort for developers.

The relatively low success rate of AdbGPT can be attributed to its reliance on extracting S2R entities from bug reports and strictly following these S2R entities step by step. However, most bug reports omit necessary initialization steps (e.g., permissions, onboarding), and since our experiments did not provide manual scripts to handle them, AdbGPT exhibited limited effectiveness. In contrast, LTGDroid, ReBL, and their variants do not depend on S2R extraction. Instead, they plan UI actions directly based on the entire bug report and contextual information such as exploration history. This allows the LLM to reason about and generate the necessary UI actions to handle simple initialization steps that may be missing from the report.

From Table~\ref{tab:evaluation-results}, we observe that ReBL-visual achieves higher success rates on both crash and non-crash bugs compared to its original version ReBL. By analyzing the additional cases successfully reproduced by ReBL-visual, we identify two main reasons. First, for some apps, the view hierarchy alone does not provide sufficient semantic cues to infer the correct UI actions, particularly when developers neglect accessibility design. In such scenarios, the visual modality introduced by UI screenshots enables ReBL-visual to make more accurate predictions. Second, non-crash bug reports often describe error symptoms in terms of visual manifestations. Since these descriptions are difficult to align with the view hierarchy, ReBL struggles to determine whether the reproduction is successful, whereas ReBL-visual benefits from the additional visual context. These findings highlight the necessity of adopting multimodal models and incorporating multimodal information for more robust bug reproduction.

We further analyze the performance differences between crash and non-crash bugs.
% , the success rates for crash bugs are 11.76\%, 67.98\%, 70.59\% and 88.82\% for AdbGPT, ReBL, ReBL-visual and LTGDroid, respectively. For non-crash bugs, the corresponding success rates are 16.67\%, 38.88\%, 52.79\% and 84.71\%.
Overall, most methods perform worse on non-crash bugs than on crash bugs, with AdbGPT and ReBL showing particularly limited capability in reproducing non-crash bugs.
Based on our observations from the non-crash bug reports, reproducing non-crash bugs often appears to be more challenging. This may be because they generally require more UI interactions to trigger, and correctly identifying each necessary action can be more difficult. This observation aligns well with our experimental results, as LLMs require sufficient contextual information to infer the correct reproduction UI actions. Specifically, LTGDroid leverages visual information along with a pre-assessment mechanism to provide the LLM with the most comprehensive contextual information; ReBL-visual provides only visual information of the UI but lacks pre-assessment context; ReBL lacks UI visual information compared to ReBL-visual; and AdbGPT relies on the least information, missing both non-S2R information from the bug reports and feedback information. Consequently, LTGDroid achieves a significantly higher success rate on non-crash bugs, further illustrating the effectiveness of its design in enhancing the overall bug reproduction capability.

% In terms of efficiency, LTGDroid required more UI actions on average, performing 23.00, 16.12, and 17.71 additional actions compared to the three baselines respectively. This is mainly due to its pre-assessment mechanism, which actively explores more potentially useful UI actions for reproduction, thereby introducing extra overhead. Although this mechanism slightly reduces efficiency, it yields a substantially higher reproduction success rate. Notably, for successful bug reproduction cases, LTGDroid indeed performs more actions on average.
% In contrast, existing methods often reach the predefined maximum number of actions in unsuccessful cases, as they repeatedly attempt UI actions without effectively progressing toward reproduction, which may cause the method to spend additional actions without making progress until this maximum is reached.

In terms of efficiency, LTGDroid required more resource consumption. This can be explained by two main factors. First, LTGDroid is able to successfully reproduce bugs with the highest interaction complexity, and reproducing such complex bugs inherently requires more resource consumption. Second, LTGDroid employs a pre-assessment mechanism that actively explores more potentially useful UI actions for reproduction, which introduces additional overhead. In future work, we plan to improve the efficiency of LTGDroid by reducing LLM token consumption and accelerating the exploration process.

% The complete results of all 75 bug reports are presented in two separate tables: Table~\ref{tab:bugs-crash-result} shows the detailed outcomes for the 51 crash bugs, whereas Table~\ref{tab:bugs-non-crash-result} summarizes the results for the 24 non-crash bugs.

% \begin{mybox}
%     \textbf{Answer of RQ3:} LTGDroid achieves significantly higher bug reproduction success rates than the baselines (AdbGPT, ReBL, and ReBL-visual), although it requires more UI actions on average due to its comprehensive exploration strategy. This trade-off results in superior effectiveness at the cost of slightly reduced efficiency.
% \end{mybox}

\begin{mybox}
    \textbf{Answer of RQ3:} LTGDroid achieves higher bug reproduction success rates than the baseline methods (AdbGPT, ReBL, and ReBL-visual) across both crash and non-crash reports, and can effectively handle complex bugs involving more UI interactions. Although its exploration strategies introduce additional resource overhead, they provide more complete contextual information for the LLM, leading to more accurate and reliable reproduction results. In future work, we plan to further reduce token consumption and accelerate the exploration process to enhance overall efficiency.
\end{mybox}

\section{Threats to Validity}

The main threat to the external validity of this study lies in the representativeness of the dataset used for evaluation. To reduce this risk, about two-thirds of the bug reports were collected from prior related studies to ensure comparability with existing work. In addition, considering that the datasets used in those studies may suffer from obsolescence, we further collected bug reports released after 2024 from the applications included in previous studies as well as several popular open-source apps, thereby improving the coverage and realism of our dataset.

Internal validity mainly concerns two aspects. First, evaluating whether a bug is successfully reproduced may introduce human judgment errors.
To improve reliability, the three authors conducted a cross-checking procedure in which each execution result was independently examined by two authors, and any disagreement was resolved through discussion with the third author. This collaborative evaluation approach helps reduce individual bias and enhances the objectivity of the assessment results.
Second, the inherent randomness of large language models may lead to variations across different experimental runs. To mitigate this issue, each experiment was executed three times, and we report the average results across all runs for all metrics to reduce the impact of random variations.

\section{Related Work}

\subsection{Automated Bug Reproduction}

Numerous approaches~\cite{ReproBot, ScopeDroid, Yakusu, ReCDroid,ReCDroid+,DroidBot, Roam, AdbGPT, ReBL} have been proposed on studying automated bug reproduction. Specifically, early research primarily relied on NLP techniques to extract S2R from bug reports and match them with UI actions or exploration paths of the application. These methods often integrated program analysis or search strategies to compensate for the incompleteness of information in bug reports.
For example, Fazzini et al. proposed Yakusu~\cite{Yakusu}, combining program analysis and NLP to extract reproduction steps from bug reports and map them to UI actions. Zhao et al. proposed ReCDroid~\cite{ReCDroid} and ReCDroid+~\cite{ReCDroid+}, which integrate NLP, deep learning, and dynamic GUI exploration to automatically generate event sequences for reproducing reported crashes. Zhang et al. introduced ReproBot~\cite{ReproBot}, which combines NLP analysis with reinforcement learning and leverages search strategies to generate successful reproduction steps. Huang et al. proposed ScopeDroid~\cite{ScopeDroid}, which builds a state transition graph (STG) based on DroidBot~\cite{DroidBot} and leverages a multimodal neural model to match bug report steps with UI states and actions, while dynamically expanding the STG to improve reproduction effectiveness. In addition, Zhang et al. proposed Roam~\cite{Roam}, which pre-constructs a comprehensive STG and matches potential reproduction paths within it for automated bug reproduction.

With the rise of LLMs, an increasing number of studies have explored leveraging their natural language understanding and reasoning capabilities to assist bug reproduction. 
Feng et al. proposed AdbGPT~\cite{AdbGPT}, which follows the conventional two-stage paradigm of S2R entity extraction and matching, similar to the aforementioned approaches. However, this paradigm has a fundamental limitation: if the S2R description does not provide a complete sequence starting from the initialization step, the reproduction cannot succeed. Our results demonstrate that AdbGPT, constrained by this conventional paradigm, exhibits limited effectiveness.
Wang et al. presented ReBL~\cite{ReBL}, which employs a feedback-driven mechanism that jointly considers bug reports, app states, and reproduction history to guide the LLM in more robust bug reproduction. However, ReBL still relies heavily on the LLM’s intrinsic ability to interpret UI semantics, which can sometimes lead to incorrect UI action generation. Our experimental results further demonstrate that LTGDroid achieves higher effectiveness compared with ReBL.

Other studies have approached bug reproduction from videos and crash stack traces. For example, Feng et al. proposed GIFdroid~\cite{GIFdroid}, which leverages screen recording and image processing techniques to automate the reproduction of bugs. Huang et al. introduced CrashTranslator~\cite{CrashTranslator}, which targets stack traces in crash reports and employs the LLM to extract key crash-related information for reproducing crashes in the app. Zhang et al. developed BugSpot~\cite{BugSpot}, which extends the use of LLMs by extracting observable error symptoms from bug reports to assist in identifying target failures during the reproduction process.

\subsection{Large Language Models for Software Engineering}

In recent years, the application of large language models in software engineering has been rapidly emerging~\cite{LLMforSE}, demonstrating significant advantages across multiple key tasks.  
% For code generation, Mu et al. proposed the ClarifyGPT framework~\cite{ClarifyGPT}, which improves the accuracy and practicality of LLM-based code generation by detecting ambiguities in requirements and generating clarification questions.  
% Zamfirescu-Pereira et al.~\cite{BeyondCodeGeneration} introduced an approach to integrate LLMs into IDEs, enabling programmers to explore a broader design space during iterative development rather than being constrained to single-shot code generation.  
% Nunez et al. presented the AutoSafeCoder~\cite{nunez2024autosafecodermultiagentframeworksecuring}, which leverages a multi-agent collaboration mechanism combining static analysis and fuzz testing to enhance the security of LLM-based code generation.  
For test generation, Zhang et al. investigated how LLMs can automatically synthesize diverse oracle verifiers~\cite{zhang2023algosynthesizingalgorithmicprograms}.  
Xue et al. proposed the LLM4Fin~\cite{LLM4Fin}, which integrates LLMs with algorithmic techniques to automatically generate high-coverage test cases from natural language business rules for fintech software acceptance testing.  
Chen et al. introduced the ChatUniTest~\cite{ChatUniTest}, which incorporates an adaptive focus-context mechanism and a generate–verify–repair pipeline to improve both accuracy and coverage in LLM-based unit test generation. In GUI testing, Ran et al. proposed the Guardian~\cite{ranGuardianRuntimeFramework2024}, which enhances the effectiveness of LLM-based feature-driven UI testing by constraining and dynamically correcting UI action planning at runtime.  
Liu et al. developed the GPTDroid~\cite{GPTDroid}, which formulates mobile GUI testing as a question-answering task and introduces functionality-aware memory prompting, enabling LLMs to perform long-term reasoning and interaction to improve test coverage and bug detection. Overall, LLMs in software engineering remain an evolving research frontier. Our work contributes to this line of research by introducing a systematic UI action exploration and pre-assessment mechanism, providing a more robust way to apply LLMs in GUI exploration and bug reproduction.

\section{Conclusion}

In this paper, we proposed LTGDroid, a novel approach that assists LLMs in bug reproduction by enumerating and pre-assessing UI actions. Unlike existing methods that solely rely on UI semantics and contextual history, LTGDroid effectively mitigates errors caused by incorrect actions, thereby improving the overall reproduction success rate. On a benchmark dataset comprising 75 bug reports, LTGDroid achieved a success rate of 85.33\%, requiring an average of 27.50 UI actions to complete each task, which demonstrates its advantage of trading moderate efficiency overhead for higher reliability. For future work, we plan to enhance LTGDroid’s capability to track exploration progress, improve its ability to recognize manifestations of non-crash bugs, and design more efficient path exploration mechanisms, with the goal of further increasing its practicality and applicability in complex scenarios.

\section{Data Availability}

The implementation of LTGDroid, together with the dataset of benchmark bug reports used in our study, are publicly available on the project website \href{https://github.com/N3onFlux/LTGDroid}{https://github.com/N3onFlux/LTGDroid}.

\vfill

\bibliographystyle{IEEEtran}
\bibliography{reference}

@misc{LTGDroid,
  author = {Github},
  title  = {LTGDroid},
  url    = {https://github.com/N3onFlux/LTGDroid},
  year   = {2025}
}

@misc{APPLAUSE,
  author = {David Bolton},
  title  = {88\% Of People Will Abandon An App Because Of Bugs},
  url    = {https://www.applause.com/blog/app-abandonment-bug-testing},
  year   = {2025}
}

@misc{GithubAPI,
  author = {GitHub},
  title  = {GitHub REST API documentation},
  url    = {https://docs.github.com/en/rest},
  year   = {2022}
}

@misc{AmazeFileManager,
  author = {Github},
  title  = {AmazeFileManager},
  url    = {https://github.com/TeamAmaze/AmazeFileManager},
  year   = {2025}
}

@misc{ChatGPT,
  author = {OpenAI},
  title  = {Introducing ChatGPT},
  url    = {https://openai.com/index/chatgpt/},
  year   = {2025}
}

@misc{GPT-4,
  author = {OpenAI},
  title  = {GPT-4},
  url    = {https://openai.com/index/gpt-4-research/},
  year   = {2025}
}

@misc{githubissuetracker,
  author = {GitHub},
  title  = {GitHub Issue Tracker},
  url    = {https://github.com/issues/},
  year   = {2025}
}

@misc{googlecode,
  author = {Google},
  title  = {Google Code Issue Tracker},
  url    = {https://code.google.com/archive/},
  year   = {2025}
}

@misc{PaLM,
  title = {{{PaLM}}: {{Scaling Language Modeling}} with {{Pathways}}},
  shorttitle = {{{PaLM}}},
  author = {Chowdhery, Aakanksha and Narang, Sharan and Devlin, Jacob and Bosma, Maarten and Mishra, Gaurav and Roberts, Adam and Barham, Paul and Chung, Hyung Won and Sutton, Charles and Gehrmann, Sebastian and Schuh, Parker and Shi, Kensen and Tsvyashchenko, Sasha and Maynez, Joshua and Rao, Abhishek and Barnes, Parker and Tay, Yi and Shazeer, Noam and Prabhakaran, Vinodkumar and Reif, Emily and Du, Nan and Hutchinson, Ben and Pope, Reiner and Bradbury, James and Austin, Jacob and Isard, Michael and {Gur-Ari}, Guy and Yin, Pengcheng and Duke, Toju and Levskaya, Anselm and Ghemawat, Sanjay and Dev, Sunipa and Michalewski, Henryk and Garcia, Xavier and Misra, Vedant and Robinson, Kevin and Fedus, Liam and Zhou, Denny and Ippolito, Daphne and Luan, David and Lim, Hyeontaek and Zoph, Barret and Spiridonov, Alexander and Sepassi, Ryan and Dohan, David and Agrawal, Shivani and Omernick, Mark and Dai, Andrew M. and Pillai, Thanumalayan Sankaranarayana and Pellat, Marie and Lewkowycz, Aitor and Moreira, Erica and Child, Rewon and Polozov, Oleksandr and Lee, Katherine and Zhou, Zongwei and Wang, Xuezhi and Saeta, Brennan and Diaz, Mark and Firat, Orhan and Catasta, Michele and Wei, Jason and {Meier-Hellstern}, Kathy and Eck, Douglas and Dean, Jeff and Petrov, Slav and Fiedel, Noah},
  year = {2022},
  month = oct,
  number = {arXiv:2204.02311},
  eprint = {2204.02311},
  primaryclass = {cs},
  publisher = {arXiv},
  doi = {10.48550/arXiv.2204.02311},
  urldate = {2025-08-23},
  archiveprefix = {arXiv}
}

@misc{LLaMA,
  title = {{{LLaMA}}: {{Open}} and {{Efficient Foundation Language Models}}},
  shorttitle = {{{LLaMA}}},
  author = {Touvron, Hugo and Lavril, Thibaut and Izacard, Gautier and Martinet, Xavier and Lachaux, Marie-Anne and Lacroix, Timoth{\'e}e and Rozi{\`e}re, Baptiste and Goyal, Naman and Hambro, Eric and Azhar, Faisal and Rodriguez, Aurelien and Joulin, Armand and Grave, Edouard and Lample, Guillaume},
  year = {2023},
  month = feb,
  number = {arXiv:2302.13971},
  eprint = {2302.13971},
  primaryclass = {cs},
  publisher = {arXiv},
  doi = {10.48550/arXiv.2302.13971},
  urldate = {2025-08-23},
  archiveprefix = {arXiv}
}

@inproceedings{Yakusu,
  title = {Automatically Translating Bug Reports into Test Cases for Mobile Apps},
  booktitle = {Proceedings of the 27th {{ACM SIGSOFT International Symposium}} on {{Software Testing}} and {{Analysis}}},
  author = {Fazzini, Mattia and Prammer, Martin and {d'Amorim}, Marcelo and Orso, Alessandro},
  year = {2018},
  month = jul,
  series = {{{ISSTA}} 2018},
  pages = {141--152},
  publisher = {Association for Computing Machinery},
  address = {New York, NY, USA},
  doi = {10.1145/3213846.3213869},
  urldate = {2025-08-17},
  isbn = {978-1-4503-5699-2},
  langid = {american}
}

@inproceedings{Andror2,
  title = {Andror2: {{A Dataset}} of {{Manually-Reproduced Bug Reports}} for {{Android}} Apps},
  shorttitle = {Andror2},
  booktitle = {2021 {{IEEE}}/{{ACM}} 18th {{International Conference}} on {{Mining Software Repositories}} ({{MSR}})},
  author = {Wendland, Tyler and Sun, Jingyang and Mahmud, Junayed and Mansur, S. M. Hasan and Huang, Steven and Moran, Kevin and Rubin, Julia and Fazzini, Mattia},
  year = {2021},
  month = may,
  pages = {600--604},
  issn = {2574-3864},
  doi = {10.1109/MSR52588.2021.00082},
  urldate = {2025-08-08},
  publisher = {{IEEE}},
  address = {Madrid, Spain},
}

@inproceedings{Themis,
  title = {Benchmarking Automated {{GUI}} Testing for {{Android}} against Real-World Bugs},
  booktitle = {Proceedings of the 29th {{ACM Joint Meeting}} on {{European Software Engineering Conference}} and {{Symposium}} on the {{Foundations}} of {{Software Engineering}}},
  author = {Su, Ting and Wang, Jue and Su, Zhendong},
  year = {2021},
  month = aug,
  series = {{{ESEC}}/{{FSE}} 2021},
  pages = {119--130},
  publisher = {Association for Computing Machinery},
  address = {New York, NY, USA},
  doi = {10.1145/3468264.3468620},
  urldate = {2025-08-08},
  isbn = {978-1-4503-8562-6}
}

@inproceedings{ReCDroid,
  title = {{{ReCDroid}}: {{Automatically Reproducing Android Application Crashes}} from {{Bug Reports}}},
  shorttitle = {{{ReCDroid}}},
  booktitle = {2019 {{IEEE}}/{{ACM}} 41st {{International Conference}} on {{Software Engineering}} ({{ICSE}})},
  author = {Zhao, Yu and Yu, Tingting and Su, Ting and Liu, Yang and Zheng, Wei and Zhang, Jingzhi and G.J. Halfond, William},
  year = {2019},
  publisher = {IEEE},
  address = {Montréal, QC, Canada},
  month = may,
  pages = {128--139},
  issn = {1558-1225},
  doi = {10.1109/ICSE.2019.00030},
  urldate = {2025-08-23}
}

@article{ReCDroid+,
  title = {{{ReCDroid}}+: {{Automated End-to-End Crash Reproduction}} from {{Bug Reports}} for {{Android Apps}}},
  shorttitle = {{{ReCDroid}}+},
  author = {Zhao, Yu and Su, Ting and Liu, Yang and Zheng, Wei and Wu, Xiaoxue and Kavuluru, Ramakanth and Halfond, William G. J. and Yu, Tingting},
  year = {2022},
  month = mar,
  journal = {ACM Trans. Softw. Eng. Methodol.},
  volume = {31},
  number = {3},
  pages = {36:1--36:33},
  issn = {1049-331X},
  doi = {10.1145/3488244},
  urldate = {2025-08-05},
  langid = {american}
}

@inproceedings{ReproBot,
  title = {Automatically {{Reproducing Android Bug Reports}} Using {{Natural Language Processing}} and {{Reinforcement Learning}}},
  booktitle = {Proceedings of the 32nd {{ACM SIGSOFT International Symposium}} on {{Software Testing}} and {{Analysis}}},
  author = {Zhang, Zhaoxu and Winn, Robert and Zhao, Yu and Yu, Tingting and Halfond, William G.J.},
  year = {2023},
  month = jul,
  series = {{{ISSTA}} 2023},
  pages = {411--422},
  publisher = {Association for Computing Machinery},
  address = {New York, NY, USA},
  doi = {10.1145/3597926.3598066},
  urldate = {2025-08-04},
  isbn = {979-8-4007-0221-1}
}

@inproceedings{ScopeDroid,
  title = {Context-{{Aware Bug Reproduction}} for {{Mobile Apps}}},
  booktitle = {Proceedings of the 45th {{International Conference}} on {{Software Engineering}}},
  author = {Huang, Yuchao and Wang, Junjie and Liu, Zhe and Wang, Song and Chen, Chunyang and Li, Mingyang and Wang, Qing},
  year = {2023},
  month = jul,
  series = {{{ICSE}} '23},
  pages = {2336--2348},
  publisher = {IEEE Press},
  address = {Melbourne, Victoria, Australia},
  doi = {10.1109/ICSE48619.2023.00196},
  urldate = {2025-08-04},
  isbn = {978-1-6654-5701-9},
  langid = {american},
  lccn = {A}
}

@inproceedings{AdbGPT,
  title = {Prompting {{Is All You Need}}: {{Automated Android Bug Replay}} with {{Large Language Models}}},
  shorttitle = {Prompting {{Is All You Need}}},
  booktitle = {Proceedings of the {{IEEE}}/{{ACM}} 46th {{International Conference}} on {{Software Engineering}}},
  author = {Feng, Sidong and Chen, Chunyang},
  year = {2024},
  month = feb,
  eprint = {2306.01987},
  pages = {1--13},
  publisher = {ACM},
  address = {Lisbon Portugal},
  doi = {10.1145/3597503.3608137},
  urldate = {2024-11-24},
  archiveprefix = {arXiv},
  isbn = {979-8-4007-0217-4},
  langid = {english},
  lccn = {A}
}

@inproceedings{CrashTranslator,
  title = {{{CrashTranslator}}: {{Automatically Reproducing Mobile Application Crashes Directly}} from {{Stack Trace}}},
  shorttitle = {{{CrashTranslator}}},
  booktitle = {Proceedings of the {{IEEE}}/{{ACM}} 46th {{International Conference}} on {{Software Engineering}}},
  author = {Huang, Yuchao and Wang, Junjie and Liu, Zhe and Wang, Yawen and Wang, Song and Chen, Chunyang and Hu, Yuanzhe and Wang, Qing},
  year = {2024},
  month = feb,
  series = {{{ICSE}} '24},
  pages = {1--13},
  publisher = {Association for Computing Machinery},
  address = {New York, NY, USA},
  doi = {10.1145/3597503.3623298},
  urldate = {2025-06-03},
  isbn = {979-8-4007-0217-4},
  lccn = {A}
}

@article{Roam,
  title = {Mobile {{Bug Report Reproduction}} via {{Global Search}} on the {{App UI Model}}},
  author = {Zhang, Zhaoxu and Tawsif, Fazle Mohammed and Ryu, Komei and Yu, Tingting and Halfond, William G. J.},
  year = {2024},
  month = jul,
  journal = {Reproduction Package for "Mobile Bug Report Reproduction via Global Search on the App UI Model"},
  volume = {1},
  number = {FSE},
  pages = {117:2656--117:2676},
  doi = {10.1145/3660824},
  urldate = {2025-06-04},
  langid = {american},
  lccn = {Not Found}
}

@inproceedings{ReBL,
  title = {Feedback-{{Driven Automated Whole Bug Report Reproduction}} for {{Android Apps}}},
  booktitle = {Proceedings of the 33rd {{ACM SIGSOFT International Symposium}} on {{Software Testing}} and {{Analysis}}},
  author = {Wang, Dingbang and Zhao, Yu and Feng, Sidong and Zhang, Zhaoxu and Halfond, William G. J. and Chen, Chunyang and Sun, Xiaoxia and Shi, Jiangfan and Yu, Tingting},
  year = {2024},
  month = sep,
  series = {{{ISSTA}} 2024},
  pages = {1048--1060},
  publisher = {Association for Computing Machinery},
  address = {New York, NY, USA},
  doi = {10.1145/3650212.3680341},
  urldate = {2025-08-04},
  isbn = {979-8-4007-0612-7},
  langid = {american},
  lccn = {A}
}

@article{BugSpot,
  title = {Automated {{Recognition}} of {{Buggy Behaviors}} from {{Mobile Bug Reports}}},
  author = {Zhang, Zhaoxu and Ryu, Komei and Yu, Tingting and Halfond, William G.J.},
  year = {2025},
  month = jun,
  journal = {Proc. ACM Softw. Eng.},
  volume = {2},
  number = {FSE},
  pages = {FSE100:2240--FSE100:2263},
  doi = {10.1145/3729370},
  urldate = {2025-08-05},
  lccn = {Not Found}
}

@inproceedings{DroidBot,
  title = {{{DroidBot}}: A Lightweight {{UI-guided}} Test Input Generator for {{Android}}},
  shorttitle = {{{DroidBot}}},
  booktitle = {Proceedings of the 39th {{International Conference}} on {{Software Engineering Companion}}},
  author = {Li, Yuanchun and Yang, Ziyue and Guo, Yao and Chen, Xiangqun},
  year = {2017},
  month = may,
  series = {{{ICSE-C}} '17},
  pages = {23--26},
  publisher = {IEEE Press},
  address = {Buenos Aires, Argentina},
  doi = {10.1109/ICSE-C.2017.8},
  urldate = {2024-12-01},
  isbn = {978-1-5386-1589-8},
  langid = {american},
  lccn = {A}
}

@inproceedings{GIFdroid,
  title = {{{GIFdroid}}: Automated Replay of Visual Bug Reports for {{Android}} Apps},
  shorttitle = {{{GIFdroid}}},
  booktitle = {Proceedings of the 44th {{International Conference}} on {{Software Engineering}}},
  author = {Feng, Sidong and Chen, Chunyang},
  year = {2022},
  month = jul,
  series = {{{ICSE}} '22},
  pages = {1045--1057},
  publisher = {Association for Computing Machinery},
  address = {New York, NY, USA},
  doi = {10.1145/3510003.3510048},
  urldate = {2025-06-03},
  isbn = {978-1-4503-9221-1},
  langid = {american},
  lccn = {A}
}

@inproceedings{wei2022chain,
author = {Wei, Jason and Wang, Xuezhi and Schuurmans, Dale and Bosma, Maarten and Ichter, Brian and Xia, Fei and Chi, Ed H. and Le, Quoc V. and Zhou, Denny},
title = {Chain-of-thought prompting elicits reasoning in large language models},
year = {2022},
isbn = {9781713871088},
publisher = {Curran Associates Inc.},
address = {Red Hook, NY, USA},
booktitle = {Proceedings of the 36th International Conference on Neural Information Processing Systems},
articleno = {1800},
numpages = {14},
location = {New Orleans, LA, USA},
series = {NIPS '22}
}

@misc{lyu2023faithful,
  title = {Faithful {{Chain-of-Thought Reasoning}}},
  author = {Lyu, Qing and Havaldar, Shreya and Stein, Adam and Zhang, Li and Rao, Delip and Wong, Eric and Apidianaki, Marianna and {Callison-Burch}, Chris},
  year = {2023},
  month = sep,
  number = {arXiv:2301.13379},
  eprint = {2301.13379},
  primaryclass = {cs},
  publisher = {arXiv},
  doi = {10.48550/arXiv.2301.13379},
  urldate = {2025-08-25},
  archiveprefix = {arXiv}
}

@inproceedings{kojima2022large,
author = {Kojima, Takeshi and Gu, Shixiang Shane and Reid, Machel and Matsuo, Yutaka and Iwasawa, Yusuke},
title = {Large language models are zero-shot reasoners},
year = {2022},
isbn = {9781713871088},
publisher = {Curran Associates Inc.},
address = {Red Hook, NY, USA},
booktitle = {Proceedings of the 36th International Conference on Neural Information Processing Systems},
articleno = {1613},
numpages = {15},
location = {New Orleans, LA, USA},
series = {NIPS '22}
}

@misc{zhang2023algosynthesizingalgorithmicprograms,
      title={ALGO: Synthesizing Algorithmic Programs with LLM-Generated Oracle Verifiers}, 
      author={Kexun Zhang and Danqing Wang and Jingtao Xia and William Yang Wang and Lei Li},
      year={2023},
      eprint={2305.14591},
      archivePrefix={arXiv},
      primaryClass={cs.CL},
      url={https://arxiv.org/abs/2305.14591}, 
}

@article{LLMforSE,
author = {Hou, Xinyi and Zhao, Yanjie and Liu, Yue and Yang, Zhou and Wang, Kailong and Li, Li and Luo, Xiapu and Lo, David and Grundy, John and Wang, Haoyu},
title = {Large Language Models for Software Engineering: A Systematic Literature Review},
year = {2024},
issue_date = {November 2024},
publisher = {Association for Computing Machinery},
address = {New York, NY, USA},
volume = {33},
number = {8},
issn = {1049-331X},
url = {https://doi.org/10.1145/3695988},
doi = {10.1145/3695988},
journal = {ACM Trans. Softw. Eng. Methodol.},
month = dec,
articleno = {220},
numpages = {79}
}

@inproceedings{LLM4Fin,
author = {Xue, Zhiyi and Li, Liangguo and Tian, Senyue and Chen, Xiaohong and Li, Pingping and Chen, Liangyu and Jiang, Tingting and Zhang, Min},
title = {LLM4Fin: Fully Automating LLM-Powered Test Case Generation for FinTech Software Acceptance Testing},
year = {2024},
isbn = {9798400706127},
publisher = {Association for Computing Machinery},
address = {New York, NY, USA},
url = {https://doi.org/10.1145/3650212.3680388},
doi = {10.1145/3650212.3680388},
booktitle = {Proceedings of the 33rd ACM SIGSOFT International Symposium on Software Testing and Analysis},
pages = {1643–1655},
numpages = {13},
location = {Vienna, Austria},
series = {ISSTA 2024}
}

@inproceedings{ChatUniTest,
author = {Chen, Yinghao and Hu, Zehao and Zhi, Chen and Han, Junxiao and Deng, Shuiguang and Yin, Jianwei},
title = {ChatUniTest: A Framework for LLM-Based Test Generation},
year = {2024},
isbn = {9798400706585},
publisher = {Association for Computing Machinery},
address = {New York, NY, USA},
url = {https://doi.org/10.1145/3663529.3663801},
doi = {10.1145/3663529.3663801},
booktitle = {Companion Proceedings of the 32nd ACM International Conference on the Foundations of Software Engineering},
pages = {572–576},
numpages = {5},
location = {Porto de Galinhas, Brazil},
series = {FSE 2024}
}

@inproceedings{ranGuardianRuntimeFramework2024,
  title = {Guardian: {{A Runtime Framework}} for {{LLM-Based UI Exploration}}},
  shorttitle = {Guardian},
  booktitle = {Proceedings of the 33rd {{ACM SIGSOFT International Symposium}} on {{Software Testing}} and {{Analysis}}},
  author = {Ran, Dezhi and Wang, Hao and Song, Zihe and Wu, Mengzhou and Cao, Yuan and Zhang, Ying and Yang, Wei and Xie, Tao},
  year = {2024},
  month = sep,
  series = {{{ISSTA}} 2024},
  pages = {958--970},
  publisher = {Association for Computing Machinery},
  address = {New York, NY, USA},
  doi = {10.1145/3650212.3680334},
  urldate = {2024-10-25},
  isbn = {979-8-4007-0612-7},
  langid = {american},
  lccn = {A}
}

@inproceedings{GPTDroid,
author = {Liu, Zhe and Chen, Chunyang and Wang, Junjie and Chen, Mengzhuo and Wu, Boyu and Che, Xing and Wang, Dandan and Wang, Qing},
title = {Make LLM a Testing Expert: Bringing Human-like Interaction to Mobile GUI Testing via Functionality-aware Decisions},
year = {2024},
isbn = {9798400702174},
publisher = {Association for Computing Machinery},
address = {New York, NY, USA},
url = {https://doi.org/10.1145/3597503.3639180},
doi = {10.1145/3597503.3639180},
booktitle = {Proceedings of the IEEE/ACM 46th International Conference on Software Engineering},
articleno = {100},
numpages = {13},
location = {Lisbon, Portugal},
series = {ICSE '24}
}

\end{document}